\documentclass[journal=jacsat,manuscript=article,layout=twocolumn,6pt]{achemso} 
\setkeys{acs}{articletitle=true,etalmode=truncate,maxauthors=0}
\usepackage{amsmath}
\usepackage{color}
\usepackage[font=footnotesize]{caption}
\usepackage{gensymb}
\usepackage{verbatim}
\author{Robert M. Elder}
\affiliation[USARL]{Polymers Branch, U.S. Army Research Laboratory, Aberdeen Proving Ground, Maryland 21005, USA}
\alsoaffiliation[BA]{Bennett Aerospace, Inc., Cary, North Carolina 27518, USA}
\alsoaffiliation[USFDA]{Present address: Center for Devices and Radiological Health, U.S. Food and Drug Administration, Silver Spring, Maryland 20903, USA}
\author{Alessio Zaccone}
\affiliation[UoM]{Department of Physics ``A. Pontremoli'', University of Milan, via Celoria 16, 20133 Milan, Italy}
\alsoaffiliation[UC]{Department of Chemical Engineering and Biotechnology, University of Cambridge, CB3 0AS Cambridge, U.K.}
\author{Timothy W. Sirk}
\email{timothy.w.sirk.civ@mail.mil}
\affiliation[USARL]{Polymers Branch, U.S. Army Research Laboratory, Aberdeen Proving Ground, Maryland 21005, USA}

\title[\texttt{achemso} demonstration]{Identifying non-affine softening modes in glassy polymer networks: A pathway to chemical design}

\begin{document}
\footnotesize 
\begin{abstract}
Using molecular simulations and theory, we develop an explicit mapping of the contribution of molecular relaxation modes in glassy thermosets to the shear modulus, where the relaxations were tuned by altering the polarity of side groups. Specifically, motions at the domain, segmental, monomer, and atomic levels are taken from molecular dynamics snapshots and directly linked with the viscoelasticity through a framework based in the lattice dynamics of amorphous solids. This unique approach provides direct insight into the roles of chemical groups in the stress response, including the timescale and spatial extent of relaxations during mechanics. Two thermoset networks with differing concentrations of polar side groups were examined, dicyclopentadiene (DCPD) and 5-norbornene-2-methanol (NBOH). A machine learning method is found to be effective for quantifying large-scale correlated motions while more local chemical relaxations are readily identified by direct inspection. The approach is broadly applicable and enables rapid predictions of the frequency-dependent modulus for any glass.
\end{abstract}
\newline
\noindent\rule{\linewidth}{1pt}
Thermoset polymers are found in a number of applications that require a combination of high elastic modulus, high yield strength, and high toughness. Unfortunately, high modulus and other characteristics of good structural performance are often accompanied by brittle behavior and poor fracture toughness. The primary method of achieving a balance of these properties is the addition of plasticizers or co-monomers that impart ductility to a strong, stiff matrix material. An alternative approach is to incrementally increase the modulus and yield strength of an inherently ductile resin through the addition of fillers or chemical functionalization. In this second approach, it is desirable to make only minimal changes to the composition of the resin so as to maintain a ductile response. Therefore, it is valuable to isolate the underlying chemical mechanisms that are responsible for the mechanical enhancements.

\begin{figure}[htb]
    \includegraphics[width=\linewidth]{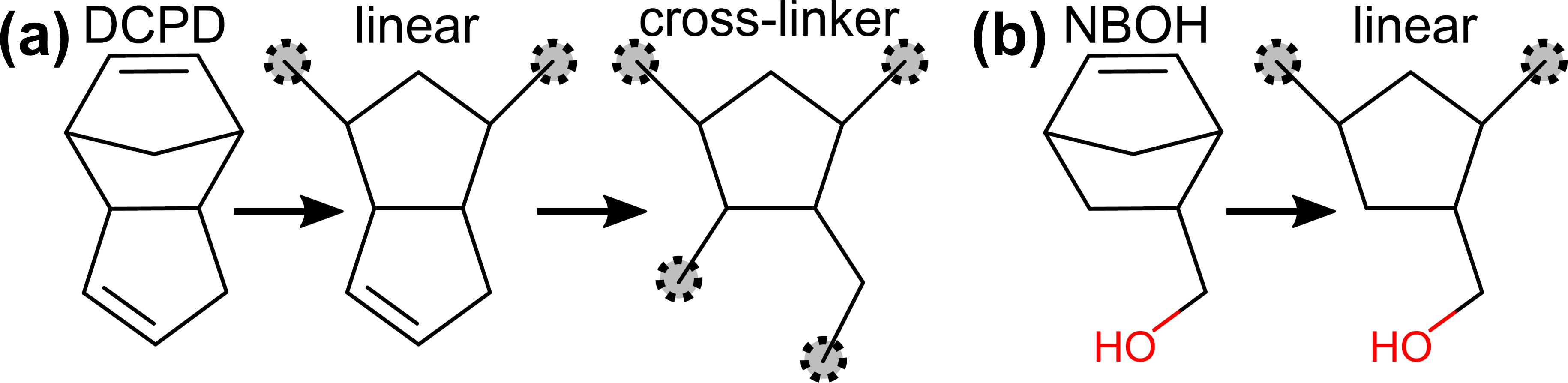}
    \caption{Chemistries studied in this work. (a) DCPD and (b) NBOH monomers can undergo one ring-opening to form linear polymers. DCPD can also undergo a second ring opening to form tetrafunctional cross-links, forming a network. Circled carbons are joined by double bonds in the network structures. See Fig. S1 for additional detail.}
    \label{fig:chem}
\end{figure}

Molecular simulations provide a potentially powerful method to understand these microscopic contributions. Previous simulations have advanced our understanding of mechanics through decompositions of the stress tensor inspired by spatial, chemical, and physical interactions (e.g., see Refs \citenum{gao1994,yoshimoto2004, vanegas2014}). However, in such decompositions, a single atom or group simultaneously participates in multiple relaxations that occur on widely different timescales. Here, we provide a new perspective by isolating the influence of specific relaxation modes in a candidate ductile matrix material, poly(dicyclopentadiene) (pDCPD), through the use of non-affine lattice dynamics (NALD) relationships \cite{lemaitre2006,zaccone2011,cui2017}.

pDCPD has emerged as a possible alternative to epoxy resins due to its ability to maintain both a high toughness and a glass transition ($\sim$140 $\degree$C) comparable to brittle structural epoxies.\cite{knorr_2016_comp_sci_tech} A unique aspect of pDCPD is that it lacks strong non-covalent interactions that act as reversible crosslinks. Here, we add transient crosslinks by substitution of hydroxyl groups in the pDCPD matrix (Fig. \ref{fig:chem}), and demonstrate the NALD relations as a promising tool to quantify and design modal contributions that soften or strengthen the quasi-static stress response.

\begin{figure}[t!]
    \includegraphics[width=1.0\linewidth]{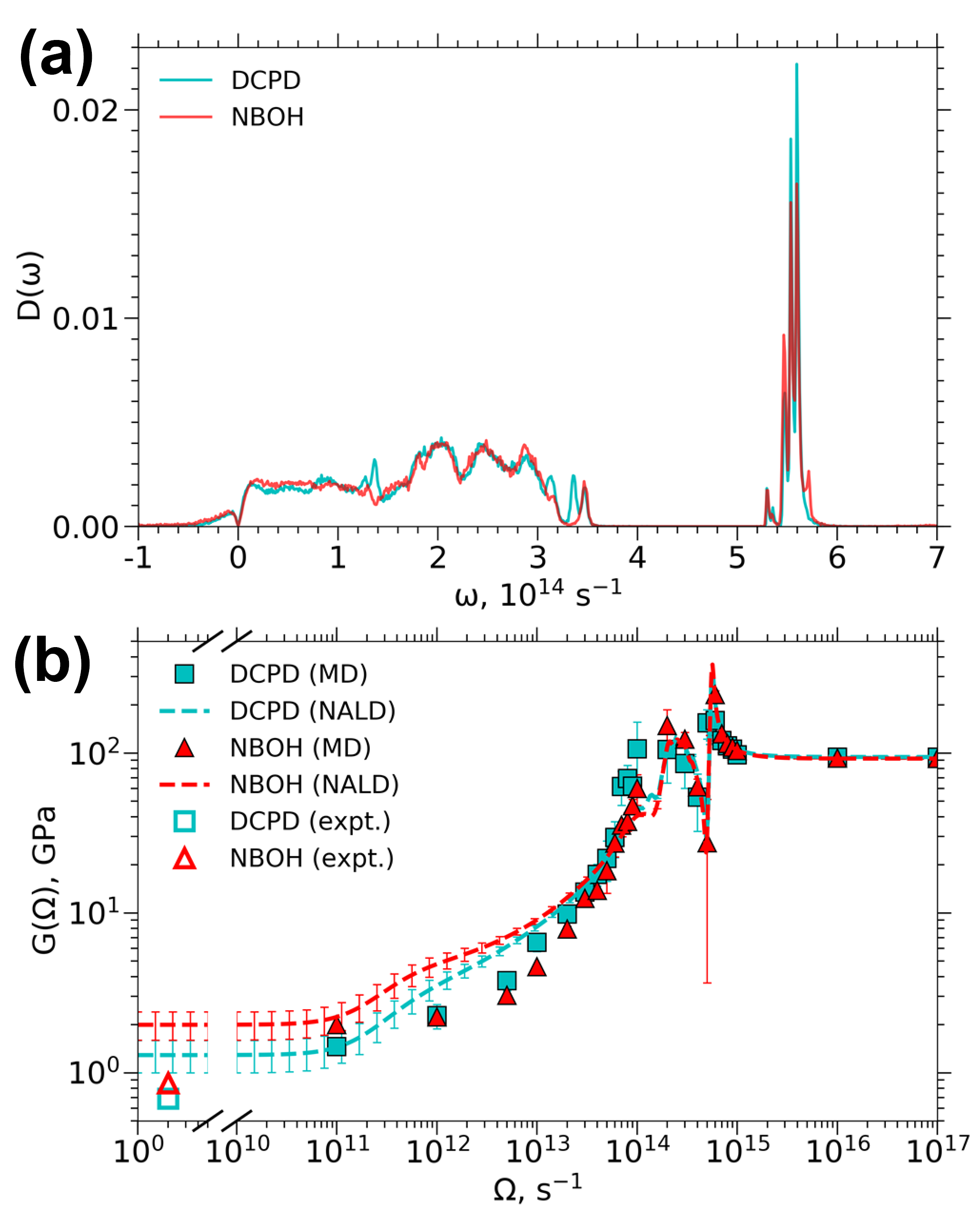}
    \caption{(a) The vibrational density of states $D(\omega)$ and (b) the shear modulus $G = \left|G^*\right|$. The error bars are the standard deviation over ten replica structures.}
    \label{fig:dos-gamma}
\end{figure}


The NALD relations require an evaluation of the potential energy surface and its spatial derivatives. In recent work at the coarse-grained level, we showed that the Hessian matrix with the addition of a tension term and inclusion of the instantaneous normal modes is an effective way to account for finite temperature in glassy polymers.\cite{vlad2018} Here, we extend this approach to atomistic systems. 

The normal modes ($\omega$) and normal mode coordinates of the Hessian inform the elastic constants $C_{\alpha\beta\kappa\chi}(\Omega)$ through a normalized vibrational density of states $D(\omega)$ and the affine force field correlator $\Gamma_{\alpha\beta\kappa\chi}(\omega)$ through the relation

\begin{equation}
C_{\alpha\beta\kappa\chi}(\Omega)=C_{\alpha\beta\kappa\chi}^{\mathrm{Born}}-3\rho\int d\omega\frac{D(\omega)\Gamma_{\alpha\beta\kappa\chi}(\omega)}{m\omega^2-m\Omega^2+i\Omega\nu},
\label{eq:C}
\end{equation}

where $\Omega$ is the applied frequency. The inputs to Eq. \ref{eq:C} are calculated using equilibrated polymer configurations from molecular dynamics (MD) simulations. A brief overview of NALD\cite{lemaitre2006,zaccone2011,cui2017} is given in the Supporting Information (SI, Section S1).

Upon inclusion of all the normal modes, Eq. \ref{eq:C} provides a prediction of the quasi-static elastic constants, though in this work we focus only on the shear modulus $G$. The first term in Eq. \ref{eq:C} corresponds to the Born approximation (i.e., the affine or high-frequency limit),\cite{born1955} while the second term accounts for non-affine effects and acts as a negative correction toward the quasi-static limit. These non-affine effects during a macroscopically affine deformation are a result of asymmetry in the atomic forces, which can be expected in all amorphous materials as well as defective or non-centrosymmetric crystals.

Identifying the contributions of specific chemical groups and molecular modes to softening has substantial practical value. Within the NALD framework these contributions to the quasi-static shear modulus can be found by reformulating Eq. \ref{eq:C} as $G(\Omega \to 0) = G_A - \sum_n \Delta G_{N\!A,n}(\Omega \to 0)$, where $G_A$ is the Born modulus. The non-affine softening contributed by a range of eigenmodes, $\omega \in \left[\omega_n, \omega_n+\Delta \omega\right]$, is found using the second term of Eq. \ref{eq:C}, i.e.,

\begin{equation}
\label{eq:dgna}
    \Delta G_{N\!A,n}(\Omega)=\
    3\rho \int_{\omega_n}^{\omega_n+\Delta\omega} d\omega\frac{D(\omega)\Gamma(\omega)}{m\omega^2-m\Omega^2+i\Omega\nu}.
\end{equation}

The set of $n$ curves $\Delta G_{N\!A,n}(\Omega)$ produces a set of $n$ values $\Delta G_{N\!A,n}(\Omega \to 0)$ that we assemble into a new curve, the frequency-dependent non-affine softening $\Delta G_{N\!A}(\omega)$, which is continuous in the limit of small $\Delta \omega$. Positive values of $\Delta G_{N\!A}(\omega)$ progressively reduce the modulus from the affine limit. Thus, even high-frequency peaks in $\Delta G_{N\!A}(\omega)$ contribute to the quasi-static modulus.

After the contributions to softening are quantified using Eq. \ref{eq:dgna}, the softening can be associated with particular motions by examining the atomic contributions to each eigenvector. To this end, we visualized the atomic eigenvectors of multiple modes near each frequency of interest and manually identified representative atomic-scale motions (e.g., C-H bond stretch). Larger-scale motions were examined using a machine learning technique wherein the atomic eigenvectors are coarse-grained via clustering. The procedure is described briefly here; Complete details are available in the SI (Section S1.5). Our goal was to group atoms that are spatially close and whose eigenvectors point in similar directions, while excluding atoms that contribute weakly to the mode. Thus, a clustering distance metric was defined that incorporates the Euclidean distance, the angle between atomic eigenvectors, and the atomic eigenvector magnitude. Atoms were then clustered using the Density-Based Spatial Clustering of Applications with Noise (DBSCAN) algorithm.\cite{DBSCAN, GDBSCAN, sklearn} Using this procedure, we identified coordinated flowing motions of neighboring monomers or chain segments. The volume of these clusters was quantified as the total Voronoi volume of the constituent atoms.\cite{voro++}

We also computed a common measure of delocalization, the participation ratio,

\begin{equation}
    \label{eq:pratio}
    p(\omega_j)=\left[ N \sum_{i=1}^N \left[ \textbf{u}_i(\omega_j) \cdot \textbf{u}_i(\omega_j) \right]^2 \right]^{-1},
\end{equation}

where $\textbf{u}_i(\omega_j)$ is the $i$-th atom's components of the normalized eigenvector of the $j$-th mode.\cite{Laird1991} For an isolated atom $p=1/N$ (complete localization), while if all atoms are involved in a single mode $p=1$ (complete delocalization). 


\begin{figure}[t!]
    \includegraphics[width=1.0\linewidth]{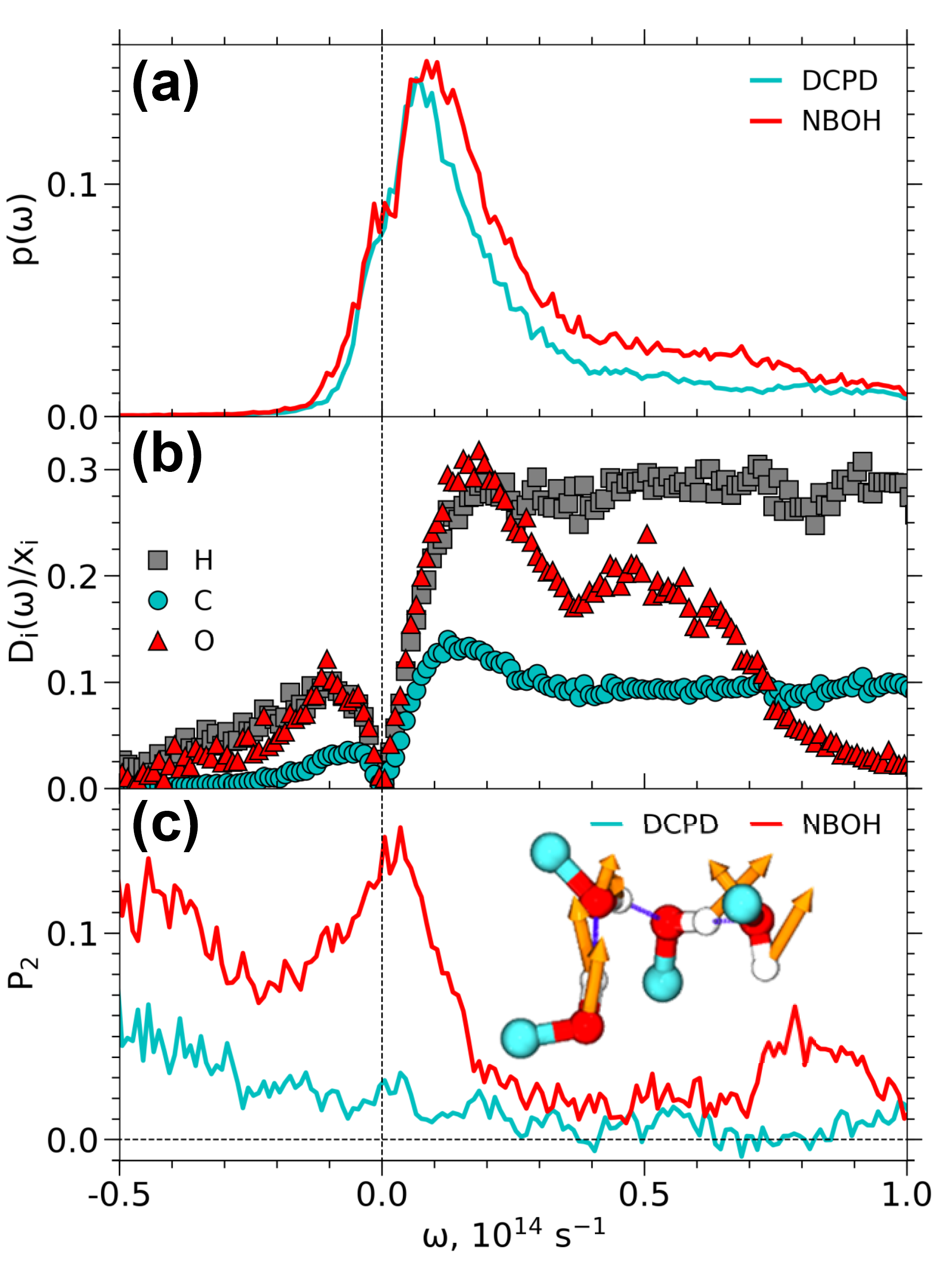}
    \caption{(a) Participation ratio $p(\omega)$. (b) Density of states $D(\omega)$ for NBOH decomposed into per-element contributions, normalized by the number concentration of each atomic element. (c) Alignment ($P_2$) of atomic eigenvectors on neighboring sidechains. Inset: H-bonded NBOH sidechains with well-aligned atomic eigenvectors.}
    \label{fig:pratio}
\end{figure}

\begin{figure*}[ht!]
    \includegraphics[width=0.8\linewidth]{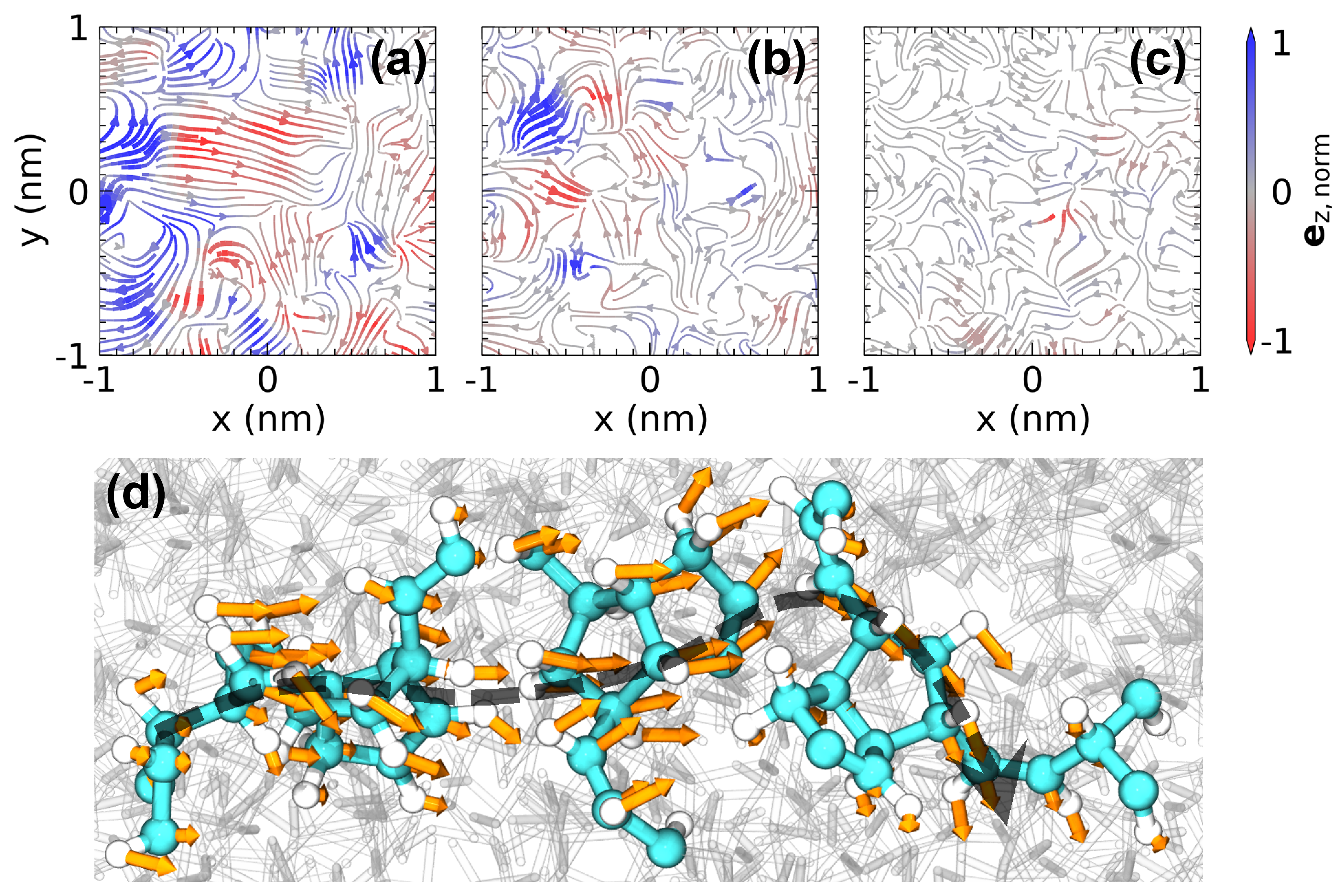}
    \caption{Streamplots of atomic eigenvectors near (a) $\omega = 10^{13}$ s$^{-1}$, (b) $\omega = 3 \times 10^{13}$ s$^{-1}$, and (c) $\omega = 10^{14}$ s$^{-1}$, where color indicates magnitude of \textit{z}-component ($\textbf{e}_{z,norm}$), and line thickness indicates total magnitude. Only atoms near the \textit{xy}-plane are included. (d) Representative snapshot of a large-scale correlated motion identified by clustering. All panels show pDCPD.}
    \label{fig:corr}
\end{figure*}

MD simulations of pDCPD and pNBOH polymer networks were carried out as in our previous work,\cite{dcpd_thermo,nanovoids,bimodal1,bimodal2} as detailed in the SI (Sections S1.3 and S1.4), and used to calculate the inputs to the non-affine expressions. We first examine the vibrational density of states $D(\omega)$. At the highest frequencies, $D(\omega)$ has several sharp peaks centered near $5.5 \times 10^{14}$ s$^{-1}$ (Fig. \ref{fig:dos-gamma}a), corresponding to hydrogen atom vibrations typical of all organic molecules. Below about $3.2 \times 10^{14}$ s$^{-1}$, a series of broad, irregularly spaced peaks appears -- the `fingerprint region' from IR spectroscopy.\cite{spechandbook} Below about $0.5 \times 10^{14}$ s$^{-1}$, $D(\omega)$ increases before falling to zero. As is customary,\cite{Stratt1995} negative frequencies are used for the unstable (imaginary) modes. These show a small peak at low negative frequencies before tapering off. We note that the affine force field correlator $\Gamma(\omega)$ shows similar patterns to $D(\omega)$ at high frequencies but rapidly decays with decreasing frequency (Fig. S2). Despite their complex character at high frequencies, both $D(\omega)$ and $\Gamma(\omega)$ demonstrate a smooth form at lower frequencies (Fig. S3) that may allow extrapolation to time- and length-scales outside the reach of the simulation data.

Next we evaluate the complex modulus $G^*$ from the NALD relations given in Eq. \ref{eq:C}. To provide a reference, non-equilibrium molecular dynamics (NEMD) simulations of mechanical spectroscopy were carried out to compute $G^*$, wherein a small sinusoidal strain field was applied to the simulation cell (see SI Section S1.4 for details). Despite several short-comings of the NEMD approach,\cite{bimodal2} the trends of $G(\Omega)$ calculated using NALD agree well with the simulated rheology results, as shown in Fig. \ref{fig:dos-gamma}b. Several significant features emerge from the atomistic and molecular modes. At the highest frequencies, $\Omega \geq 10^{15}$ s$^{-1}$, the affine modulus $G_A$ is recovered as expected. In the intermediate frequencies of $10^{11}$ s$^{-1} \leq \Omega \leq 10^{15}$ s$^{-1}$, $G$ decreases through a set of irregular peaks and troughs. Both NEMD and NALD display an overshoot near the high-frequency transitions, indicating strong mechanical resonance between the applied oscillation and the intrinsic modes of the polymers. For $\Omega \leq 10^{11}$ s$^{-1}$, $G$ approaches the low-frequency plateau. These results can be compared with experimental results at low strain rates (${<}1$ s$^{-1}$), despite the absence of spatially extended modes that exceed the size of the simulation cell. Indeed, the predicted low-frequency values for pDCPD and pNBOH ($1.3 \pm 0.3$ and $2.0 \pm 0.4$ GPa) compare reasonably well with experimental values (0.69 and 0.86 GPa).\cite{nboh} The NALD predictions capture the ordering and relative magnitude of $G$, but are elevated by a factor of $\sim$2, demonstrating that NALD provides semi-quantitative predictions of glassy mechanics. The overprediction of $G$ can be partially attributed to additional softening at lower frequencies, which are absent due to finite-size effects (see SI Sections S1.2 and S2). In these compositions, pDCPD experiences much greater softening as the frequency is reduced, where the low-frequency modulus of pDCPD is less than that of pNBOH yet the affine modulus $G_A$ of DCPD is higher. In what follows, we characterize the importance of modal contributions to the stress and identify their role in softening.

First, we examine the motions with the largest spatial extent. One approach to quantify such motions is the participation ratio $p(\omega)$, which corresponds to the fraction of atoms participating in each vibrational mode. An analysis of the pDCPD and pNBOH networks indicates that the lower-frequency bands in our simulation, near $10^{13} s^{-1}$, involve upwards of 10\% of the system. As the frequency increases, the the number of atoms sharply decreases to ${\sim}$1\% (Fig. \ref{fig:pratio}a). Interestingly, pNBOH displays a higher participation ratio in the low frequency range. To explain this result, a complementary analysis was carried out to decompose $D(\omega)$ into atomic elements. We find that the relative contribution of oxygen atoms to $D(\omega)$ is high in the same band of low frequencies (Fig. \ref{fig:pratio}b). Together, these findings suggest the influence of hydrogen bonds. We calculated the alignment $P_2 = 0.5\left(3 \langle \cos^2{\theta} \rangle -1 \right)$ of atomic eigenvectors near the sidechains (within 5 \AA), where $\theta$ is the angle between the vectors. In pDCPD, there is essentially no alignment ($P_2 \approx 0$ in Fig. \ref{fig:pratio}c), while in pNBOH, there is much better alignment when these sidechains are H-bonded (Fig. \ref{fig:pratio}c inset). In this way, H-bonds provide a mechanism to recruit neighboring atoms into a vibrational mode, and thereby increase $p(\omega)$.

\begin{figure}[tb]
    \includegraphics[width=1.0\linewidth]{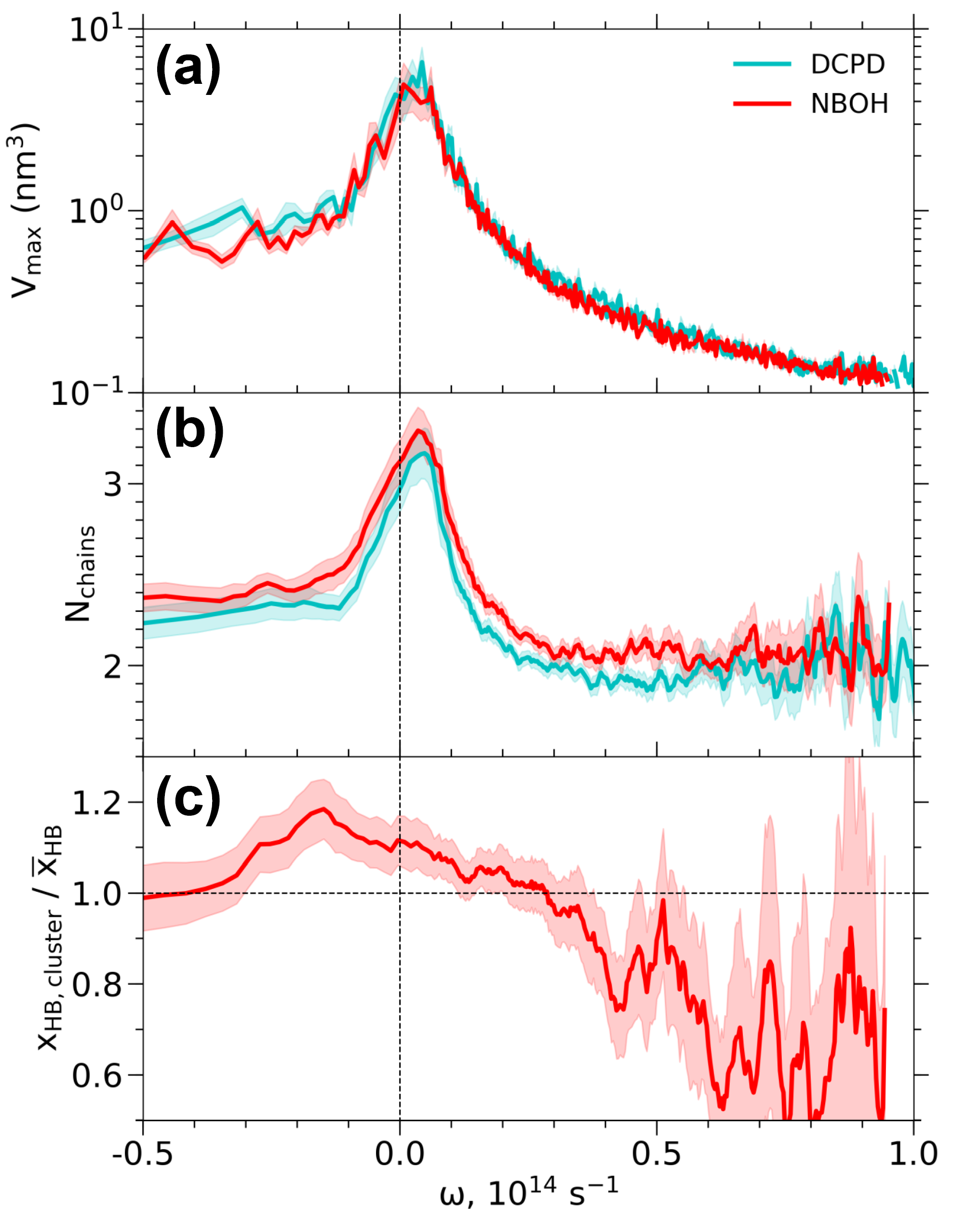}
    \caption{(a) Volume of largest clusters. (b) Average number of polymer chains per cluster. (c) Concentration of H-bonds in clusters, $x_{HB,cluster}$, normalized by the bulk concentration, $\overline{x}_{HB}$. }
    \label{fig:clust}
\end{figure}

The participation ratio only captures the fraction of atoms contributing to each mode and includes atomic eigenvectors of any direction and location. Thus, it is not a definitive measure of cooperative motion. By taking slices of the eigenvector field, it is clear that localized cooperative motions do exist at low frequencies, and their spatial extent is reduced with increasing frequency (Fig. \ref{fig:corr}a-c). To quantify these motions, we applied a machine learning algorithm to sort atoms into clusters with well-aligned atomic eigenvectors, as exemplified in Fig. \ref{fig:corr}d. This approach bears a resemblance to the coarse-grained normal mode analysis techniques that have proven so useful for structural biology (e.g., Ref \citenum{bahar2010}).

The clustered motions are restricted to spatially adjacent vectors with similar directions. At high frequencies ($\omega > 0.5 \times 10^{14}$ s$^{-1}$), these motions encompass only a few atoms, with maximum volumes $V_{max} \approx 0.1$ nm$^3$ (Fig. \ref{fig:clust}a). As $\omega$ decreases toward zero, the cooperative length scale rapidly grows to $V_{max} \approx 10$ nm$^3$, or about 20\% of the simulation volume. Apparently there is no difference between pDCPD and pNBOH in terms of volume of cooperative motion. However, clusters in pNBOH involve a slightly larger number of polymer chains (Fig. \ref{fig:clust}b), suggesting stronger inter-chain interactions. This is associated with H-bonds, as low-frequency clusters have a higher incidence of H-bonds (Fig. \ref{fig:clust}c), which echoes the results in Fig. \ref{fig:pratio}. Overall, these differences in the character of large-scale motions are a matter of degree and do not suggest a compelling explanation for the larger softening in pDCPD.

\begin{figure*}[ht!]
    \includegraphics[width=0.8\linewidth]{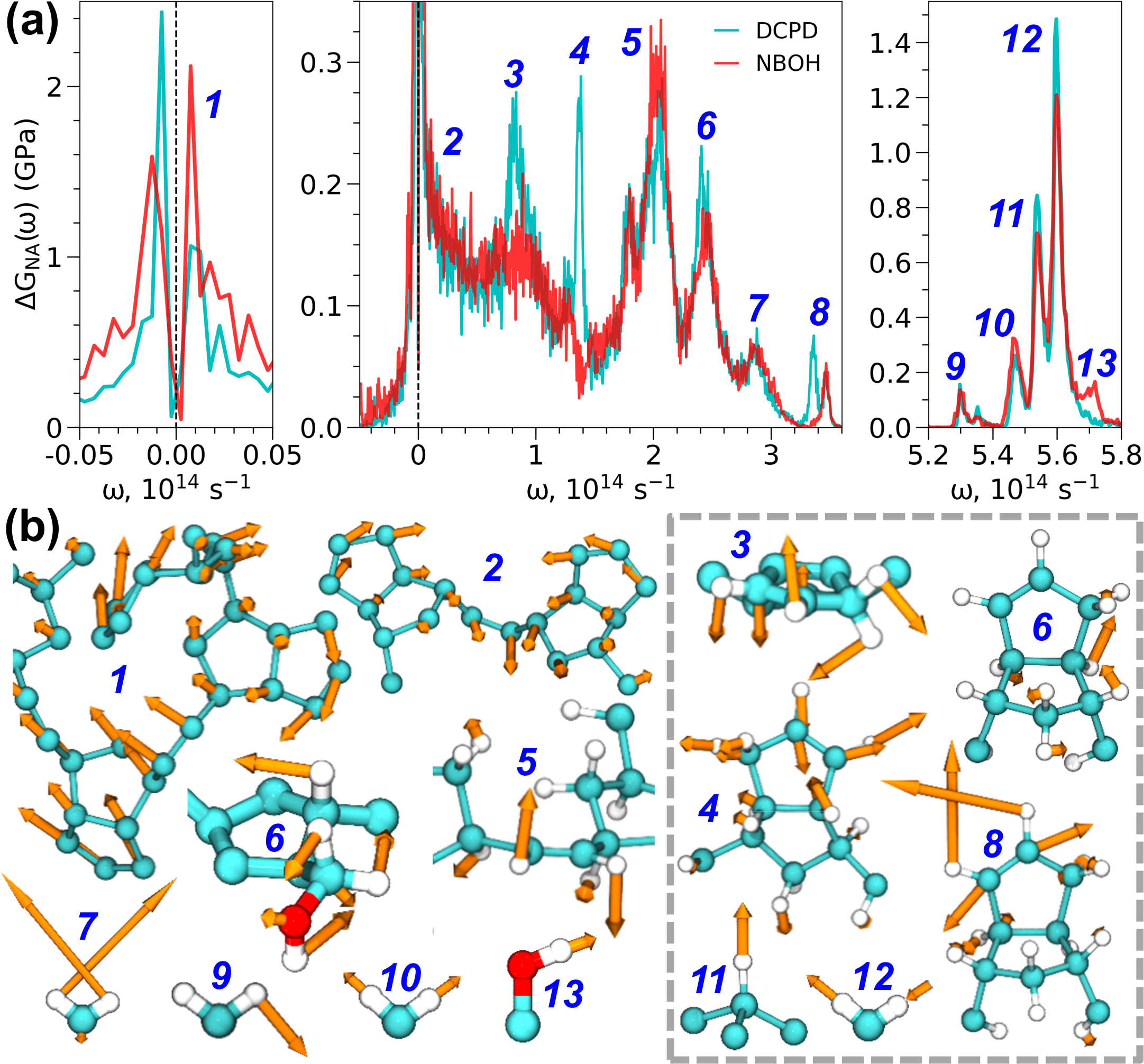}
    \caption{Non-affine softening associated with molecular motions. (a) Frequency-dependent non-affine softening, $\Delta G_{N\!A}(\omega)$, with several interesting peaks labeled. Note that the scale of the axes differs in each panel. (b) Molecular motions identified from atomic contributions to eigenmodes. Motions within the box contribute strongly to excess softening in DCPD.}
    \label{fig:gna}
\end{figure*}

Instead we turn to smaller-scale, higher-frequency modes. We identified the origins of softening using $\Delta G_{N\!A}(\omega)$, calculated from Eq. \ref{eq:dgna}, where peaks indicate large contributions to softening from the affine limit. The contributions are shown with labels in Fig. \ref{fig:gna}a. The corresponding atomic and molecular motions responsible for each peak were identified from the eigenvectors, as shown in Fig. \ref{fig:gna}b. We summarize the spectrum as follows: The highest-frequency peak (\#13, $\omega \approx 5.7 \times 10^{14}$ s$^{-1}$) originates from O-H stretching, so it only appears in pNBOH. This peak is much narrower than would be expected from spectroscopy, likely because the force field used here lacks polarizability.\cite{pande2013} Next is a large, sharp peak (\#12, $\omega \approx 5.6 \times 10^{14}$ s$^{-1}$) that corresponds to asymmetric C-H bond stretching followed by a nearby peak (\#10, ${\sim}5.5 \times 10^{14}$ s$^{-1}$) for the symmetric C-H stretch. The asymmetric and symmetric stretch are ordered correctly and the C-H frequencies agree well with experimental IR spectra (${\sim}2900$ cm$^{-1}$).\cite{spechandbook} Between these two peaks lies another (\#11, $\omega \approx 5.55 \times 10^{14}$ s$^{-1}$), which is linked to C-H stretching on sp$^3$ CH groups. At $\omega \approx 5.3 \times 10^{14}$ s$^{-1}$, the narrow peak (\#9) corresponds to C-H bond wagging in which single H atoms vibrate side-to-side, perpendicular to the bond. These hydrogen-related motions involve the stiffest, highest-frequency bonds in the system. Hence, if these bonds are effectively frozen by a high-frequency external oscillation $\Omega$, the material becomes extremely stiff and the softening $G_{N\!A}$ decreases strongly. 

Upon entering the fingerprint region, we observe a stronger distinction between pDCPD and pNBOH. First we find two sharp peaks (\#8, $\omega \approx 3.4 \times 10^{14}$ s$^{-1}$) corresponding to C=C double bond stretching. The higher-frequency peak is the backbone double bond, present in both pDCPD and pNBOH, while the lower-frequency peak is the pDCPD sidechain C=C bond, and is absent for pNBOH. Next, a low, broad peak (\#7, ${\sim}2.9 \times 10^{14}$ s$^{-1}$) corresponding to CH$_2$ scissoring and in good agreement with experimental values of ${\sim}1470$ cm$^{-1}$.\cite{spechandbook} A peak linked to CH$_2$ twisting (\#6, $\omega \approx 2.4 \times 10^{14}$ s$^{-1}$) is prominent, along with wagging of H atoms on sp$^3$ CH groups. In pNBOH, peak \#6 also includes twisting of the $-$CH$_2$OH sidechain. At the next large peak (\#5, ${\sim}2.1 \times 10^{14}$ s$^{-1}$), we find twisting of H atoms across double bonds. Next are two peaks that appear in pDCPD but not pNBOH. Peak \#4 arises from expansion and contraction (`breathing') of the pDCPD cyclopentene ring, which is absent in pNBOH.\cite{spechandbook} Peak \#3 also involves this cyclopentene ring, with twisting around the double bond coupled to CH$_2$ rotation. A similar CH$_2$ rotation occurs on the cyclopent\textit{ane} ring in both networks, similar to the `envelope flip' motion we observed in a previous study of pDCPD.\cite{dcpd_thermo} Likewise, as we approach the low-frequency limit (\#2), we find motions reminiscent of an inter-monomer twisting motion we identified previously.\cite{dcpd_thermo} Finally, at the lowest frequencies (\#1, ${\sim}10^{12}$ s$^{-1}$), we observe collective motions involving several monomers or entire chain segments. Freezing these larger-scale motions raises the stiffness ten-fold from the low-frequency limit up to $10^{13}$ s$^{-1}$ (Fig. \ref{fig:dos-gamma}b). We note that, at the lowest frequencies (Fig. \ref{fig:gna}a, left) pNBOH shows excess softening which may coincide with H-bonds given in Fig. \ref{fig:pratio} and Fig. \ref{fig:clust}. 

The peaks responsible for enhanced softening in pDCPD are summarized in Fig. \ref{fig:gna}b (gray box). At high frequencies, peaks \#11 and \#12 are larger in pDCPD. In the fingerprint region, we find larger contributions from sidechain double bonds in peak \#8, CH$_2$ twisting and sp$^3$ CH wagging in \#6, and cyclopentene ring vibrations in peaks \#3 and \#4. Overall, excess softening in pDCPD is primarily connected to the fast modes associated with its highly constrained cyclopentene sidechain, while the fast modes of the equivalent side chain in NBOH contribute less softening. Interestingly, it is clear from Figure \ref{fig:gna} that restricting other high-frequency modes in the system (e.g., twisting in peak \#5) would further prevent softening and lead to an additional increase in the modulus.


In summary, molecular simulations were applied in concert with a theory of non-affine lattice dynamics (NALD) to examine two glassy thermosets with contrasting electrostatic interactions, dicyclopentadiene (DCPD) and 5-norbornene-2-methanol (NBOH), as potential candidates for ductile matrix materials. The networks were characterized through the vibrational density of states, rate-dependent shear modulus, participation ratio, relaxation volumes of the normal modes, and modal stress contributions. We found good agreement of the shear modulus between experiments, NALD predictions, and extensive non-equilibrium molecular dynamics simulations of mechanical spectroscopy. Using the NALD approach, we found that the hydroxyl functionalization of NBOH was able to strengthen the shear modulus by reducing several key softening modes found in the high-frequency relaxations of a rigid side chain, rather than in the slower and more extended multi-chain motions. These differences were ultimately responsible for the softer modulus of DCPD in the quasi-static limit. We anticipate the NALD approach applied here can enable a more designed functionalization of glassy and semi-crystalline polymers for mechanics, in which a minimal loading of functional groups targets specific softening modes. Several paths forward are available to further enhance the analysis, such as the incorporation of the vibrational density of states from inelastic scattering experiments\cite{crupi2014} and analytical corrections for finite size effects. 

\subsection{Associated Content}

\textbf{Supporting Information.} A summary of the theory underlying non-affine lattice dynamics; extensions to the NALD method for atomistic systems; details of the chemistry (Fig. S1), simulation protocols, and analysis methods; and additional results (Figs. S2--S3).

\acknowledgement

The research reported in this document was performed in connection with contract W911QX-16-D-0014, W911NF-16-2-0091, and W911NF-19-2-0055 with ARL. The views and conclusions contained in this document are those of Bennett Aerospace, Inc. and ARL. Citation of manufacturer's or trade names does not constitute an official endorsement or approval of the use thereof. The U.S. Government is authorized to reproduce and distribute reprints for Government purposes notwithstanding any copyright notation hereon.

\renewcommand{\baselinestretch}{1}
\bibliography{na}





\end{document}